# Piezoresponse of ferroelectric films in ferroionic states: time and voltage dynamics

*Anna N. Morozovska,[1] Eugene A. Eliseev,[1,2], Nicholas V. Morozovsky[1], and Sergei V. Kalinin[3,\*]*

[1]Institute of Physics, National Academy of Sciences of Ukraine,

46, Prospekt Nauky, 03028 Kyiv, Ukraine

[2]Institute for Problems of Materials Science, National Academy of Sciences of Ukraine,

3, Krjijanovskogo, 03142 Kyiv, Ukraine

[3]Center for Nanophase Materials Science, Oak Ridge National Laboratory, Oak Ridge, Tennessee 37831, USA

## Abstract

The interplay between electrochemical surface charges and bulk ferroelectricity in thin films gives rise to a continuum of coupled ferro-ionic states. These states are exquisitely sensitive to chemical and electric conditions at the surfaces, applied voltage, and oxygen pressure. Using the analytical approach combining the Ginzburg-Landau-Devonshire description of the ferroelectricity with Langmuir adsorption isotherm for the ions at the film surface, we have studied the temperature-, time- and field- dependent polarization changes and electromechanical response of the ferro-ionic states. The responses are found to be inseparable in thermodynamic equilibrium and at low frequencies of applied voltage. The states become separable in high frequency dynamic mode due to the several orders of magnitude difference in the relaxation times of ferroelectric polarization and surface ions charge density. These studies provide an insight into dynamic behavior of nanoscale ferroelectrics with open surface exposed to different kinds of electrochemically active gaseous surrounding.

---

[\*] Corresponding author. E-mail: sergei2@ornl.gov (S.V.K.)

Within 70 years, the development of perovskite ferroelectric applications has evolved from capacitor radio-ceramics bulk units [1, 2], multifunctional piezoelectric thin sheet transducers [3, 4] and pyroelectric thin layer detectors [5] to thin film ferroelectric memories [6, 7]. From early 1970-ies to present there are growing demands to ferroelectric structures in chemotronics [ 8 ] and electrochemical sensorics [9] for design the new classes of sensing and converting devices. The first experiments [10, 11, 12, 13] have proved the interaction of the ferroelectric barium titanate, $BaTiO_3$, surface with oxygen, hydrogen and methyl-alcohol molecules resulted in the significant influence on polarization reversal, dielectric permittivity and pyroelectric currents. Electric coupling between electrochemical surface charges and bulk ferroelectricity gives rise to a chemical switching [14, 15, 16] and continuum of coupled ferro-ionic states [17, 18] in ferroelectric thin films. These states are exquisitely sensitive to chemical and electric conditions at the film surfaces [14-18]. The analytical description of coupled ferro-ionic states was developed via the combination of Ginzburg-Landau-Devonshire description of the film ferroelectric properties [19, 20, 21] with Langmuir adsorption model for the chemical reaction at the film surface [22], and corresponding phase diagrams .as a function of temperature, film thickness, and external electric potential were constructed under different stationary conditions [14-18].

Here, we explore the dynamic behavior of the ferro-ionic states in thin films, representing the polarization and electromechanical response as measured by the piezoresponse force microscopy (**PFM**) [23, 24]. We demonstrate that the piezoelectric responses in ferroelectric, ferro-ionic and non-ferroelectric electret-like ionic states are inseparable at low frequencies of applied voltage. The states become separable in high frequency dynamic regime due to the several orders of magnitude difference in the relaxation times of ferroelectric polarization and surface ions. This effect calls up with pyroelectric response under slow enough and fast enough temperature changes [25, 26]. These studies provide an insight into mesoscopic properties of ferroelectric thin films, whose surface is exposed to electrochemically active gaseous surrounding, and potentially point out experimental pathways to explore them.

To describe the coupling between ferroelectric phenomena and interfacial electrochemistry (redox reactions), we adopted the Stephenson and Highland (**SH**) approach [14, 16] and further extend it by the presence of the dielectric gap of thickness λ between the quasi-planar top electrode (e.g., flat apex of the PFM tip) and ferroelectric surface. The film thickness is *h*. The bottom electrode is flat and ideally electron conducting. Voltage *U* is applied to the tip electrode during enough long time for system to reach the thermodynamic equilibrium [**Fig. 1(a)**]. Then the voltage can be switched off [**Fig. 1(b)**]. We assume that the ion layer with the Langmuir-type charge density $\sigma(\varphi)$ [14, 16, 22] cover the ferroelectric film surface z = 0. The existence of the ion layer is necessary to provide an effective screening of the spontaneous polarization at the open surface of the ferroelectric film. Such layers can be readily formed when the film top surface was exposed to a controlled oxygen partial pressure in equilibrium with electron conducting bottom electrode. Experimentally these layers can be realized either by design (similar to electrochemical experiments), or serendipitously via adsorption from environment.

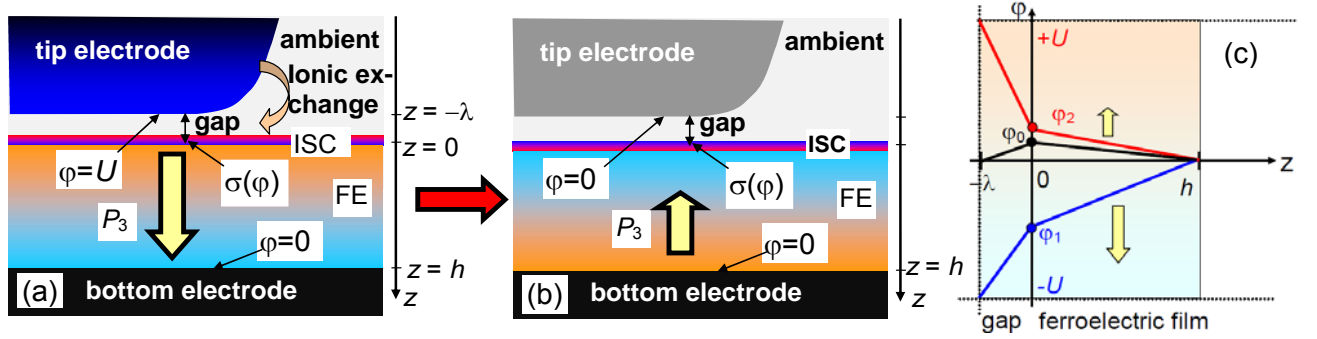

**FIGURE 1**. **(a-b)** Layout of the considered system, consisting of electron conducting bottom electrode, ferroelectric (FE) film, ion surface charge (ISC), ultra-thin gap allowing direct ion exchange with environment, and the tip electrode (from bottom to the top). **(a)** Voltage $U$ is applied to the tip electrode. **(b)** The voltage is switched off. **(c)** Schematics of electric potential z-profiles for positive (+$U$) and negative (−$U$) voltages applied to the tip (red and blue broken lines). The arrows indicate polarization direction.

The linear equation of state, $\mathbf{D} = \varepsilon_0 \varepsilon_d \mathbf{E}$, relates the electrical displacement $\mathbf{D}$ and electric field $\mathbf{E}$ in the gap. Here $\varepsilon_0$ is a universal dielectric constant, and $\varepsilon_d \sim 1$ is the relative permittivity of the physical gap media (vacuum, air or inert gas environment). The space charge density is negligibly small in a wide band-gap ferroelectric film, and so its polarization z-component depends on the inner field $\mathbf{E}$ as $P_3(\mathbf{r}, E_3) = P_3^f(\mathbf{r}, E_3) + \varepsilon_0(\varepsilon_{33}^b - 1)E_3$, where background permittivity $\varepsilon_{33}^b \leq 10$ [27]. The system of electrostatic equations in the gap and ferroelectric film, and electric boundary conditions are summarized in **Appendix A** in **Suppl.Mat.** The spatial and temporal distribution of the ferroelectric polarization $P_3$ is given by the time-dependent LGD equation [28] (see **Appendix A** in **Suppl.Mat.**) with the boundary conditions of the third kind at the film surfaces z = 0 and z = h, which include extrapolation lengths $\Lambda_+$ and $\Lambda_-$ [29, 30].

We propose a linear relaxation model for the dynamics of surface ions charge density, $\tau \frac{\partial \sigma}{\partial t} + \sigma = \sigma_0[\varphi]$. The dependence of equilibrium charge density $\sigma_0[\varphi]$ on electric potential $\varphi$ is controlled by the concentration of surface ions $\theta_i$ at the interface z = 0, $\sigma_0[\varphi] = \sum_i \frac{eZ_i \theta_i(\varphi)}{A_i}$, where $e$ is an elementary charge, $Z_i$ is the charge of the surface ions/vacancies, $\theta_i$ are relative concentration of surface ions, $A_i$ is the inverse concentration of the surface ion at saturation, at that $i = 2$ to reach the charge compensation. The surface charge is controlled by the potential φ at the interface $z = 0$ in a self-consistent manner [14, 16, 31]:

$$\theta_i(\varphi) = \left(1 + \left(\frac{p_{atm}}{p_{O2}}\right)^{1/n_i} \exp\left(\frac{\Delta G_i^{00} + eZ_i\varphi}{k_B T}\right)\right)^{-1}, \quad (1)$$

Here $p_{O2}$ is the oxygen partial pressure excess (relative to atmospheric pressure $p_{atm}$), $n_i$ is the number of surface ions created per oxygen molecule, $\Delta G_i^{00}$ is the free energy of the surface ion formation at $p_{O2} = 1$ bar and $U = 0$. Note that the developed solutions are insensitive to the specific details of the charge compensation process, and are sensitive only to the thermodynamic parameters of corresponding effective reactions [32].

Since the stabilization of single-domain polarization in ultrathin perovskite films takes place due to the chemical switching (see e.g. [14, 16, 33, 34, 35]) we consider the *single-domain* film and assume that its polarization distribution is smooth enough. For this case, we derived coupled nonlinear equations for the polarization averaged over film thickness $\langle P_3 \rangle$ and surface charge density σ [18]:

$$\Gamma \frac{\partial \langle P_3 \rangle}{\partial t} + (a_R - 2Q_{11}X_{33})\langle P_3 \rangle + a_{33}\langle P_3 \rangle^3 + a_{333}\langle P_3 \rangle^5 = E_{eff}(U, \sigma), \quad (2a)$$

$$\tau \frac{\partial \sigma}{\partial t} + \sigma = \sigma_0 \left[E_{eff}(U, \sigma)h - \frac{\lambda \langle P_3 \rangle h}{\varepsilon_0(\varepsilon_d h + \lambda \varepsilon_{33}^b)}\right]. \quad (2b)$$

In Eq. (2a) the kinetic Khalatnikov coefficient Γ is defined by phonon relaxation time [36]. Coefficient $a_R = \alpha_T(T_C - T) + \frac{g_{33}}{h}\left(\frac{1}{\Lambda_+} + \frac{1}{\Lambda_-}\right) + \frac{\lambda}{\varepsilon_0(\varepsilon_d h + \lambda \varepsilon_{33}^b)}$ is the free energy coefficient $a_3 = \alpha_T(T_C - T)$ renormalized by intrinsic gradient-correlation size effects (the term $\sim g_{33}/\Lambda$) and extrinsic depolarizing size effect (the term ~λ) [18]. $T$ is the absolute temperature, $T_C$ is Curie temperature. The term $2Q_{11}X_{33}$ originated from electrostrictive coupling [37, 38]; $Q_{11}$ is the electrostriction coefficient, $X_{33}$ is the elastic stress component. $a_{33}$, and $a_{333}$ are the coefficients of LGD potential $F(P_i, X_{ij}, \sigma)$ expansion on the higher polarization powers [19]. The effective electric field in Eqs.(2) is $E_{eff}(U, \sigma) = \frac{\lambda \sigma + \varepsilon_0 \varepsilon_d U}{\varepsilon_0(\varepsilon_d h + \lambda \varepsilon_{33}^b)}$. Electric potentials acting in the dielectric gap ($\varphi_d$) and in ferroelectric ($\varphi_f$) linearly depends on the coordinate z and overpotential $\Psi = h\left(E_{eff}(U, \sigma) - \frac{\lambda \langle P_3 \rangle}{\varepsilon_0(\varepsilon_d h + \lambda \varepsilon_{33}^b)}\right)$, namely $\varphi_d = U - \frac{z + \lambda}{\lambda}(U - \Psi)$ and $\varphi_f = (h - z)\frac{\Psi}{h}$ [**Fig. 1(c)**].

Here, we extend the analysis [18] to study the effective piezoelectric response $d_{33} = \partial \langle P_3 \rangle / \partial X_{33}$:

$$\Gamma \frac{\partial d_{33}}{\partial t} + \left( a_R + 3a_{33} \langle P_3 \rangle^2 + 5a_{333} \langle P_3 \rangle^4 \right) d_{33} = 2Q_{11} \langle P_3 \rangle, \qquad (3)$$

Expressions (1)-(3) describe the coupling between ferroelectric polarization and surface ions, and derive effective piezoelectric coefficient.

It was shown [18] that the phase diagram of the ferroelectric film covered by ions becomes more complex and the film can adopt the non-ferroelectric ionic (**NFE**), coupled ferroelectric-ionic (**CFI**) and mostly ferroelectric (**FE**) states dependent on temperature and thickness. The FE state is defined as the state with robust ferroelectric hysteresis between two absolutely stable and two unstable ferroelectric polarizations $\langle P_3 \rangle$, which have "positive" or "negative" projection at the film surface normal. The four polar states correspond to the four real roots of the static Eqs.(2a), which can exist at nonzero σ, for the film thickness $h$ higher than the critical thickness $h_{cr}(T) \approx (L_g + L_d)T_C/(T_C - T)$, and temperatures smaller than the critical one $T_{cr}(h) \approx T_C(1 - (L_g + L_d)/h)$ ($L_g$ and $L_d$ definitions are given in the last rows of **Table SI** in **Suppl.Mat.**). Positive and negative orientations of $\langle P_3 \rangle$ are energetically equivalent only at σ = 0 and $U$ = 0. At σ ≠ 0 the difference between their energies increases with the film thickness decrease. The physical origin of the $+\langle P_3 \rangle$ and $-\langle P_3 \rangle$ asymmetry is the built-in field $E_{eff}(U, \sigma)$ that is nonzero at $U$ = 0 for the different surface ion formation energies, $\Delta G_1^{00} \neq \Delta G_2^{00}$. The CFI state is characterized by one stable and one unstable polarization states. The NFE state that has no hysteresis properties in the thermodynamic limit, manifests electret-like behavior with polarization $\langle P_3 \rangle \cong E_{eff}/a_R$ induced by the built-in electric field $E_{eff}$.

**Figure 2** shows the temperature dependences of the average polarization $\langle P_3(T) \rangle$ calculated for BaTiO$_3$ films. Each curve corresponds to a defined voltage $U$. Both stable polar states (for which $\partial \langle P_3 \rangle / \partial T > 0$) and metastable ones (for which $\partial \langle P_3 \rangle / \partial T < 0$) are shown. For the films thickness less than $h_{cr}(0) \approx 11$ nm the temperature dependences $\langle P_3(T) \rangle$ are strongly asymmetric with respect to voltage [**Figs.2(a)-(b)**]. The asymmetry $\langle P_3(+U) \rangle \neq \langle P_3(-U) \rangle$ for thicker films is due to ions presence [**Figs.2(c)**], and is the most pronounced at lower temperatures. The diffuse transition of the film to the electret-like NFE state (instead of the paraelectric one) takes place at higher temperatures. Since the negative polarizations gradually reach each other with the temperature increase the diffuse boundary between NFE and CFI states is defined by the parameters of surface ionic states (such as $p_{O2}$, $A_i$, and $\Delta G_i^{00}$).

Note that even though ultra-thin epitaxial films without top electrode (or with a gap between the film surface and the top electrode) can split into domain stripes [39, 40, 41], the analysis suggests that the considered ultra-thin films are single-domain because the screening ion charge layer covers the film surface and induces the strong self-polarizing effective electric field $E_{eff}(U,\sigma)$. The field, being inversely proportional to the film thickness, is nonzero at $U=0$ for the case of different ion formation energies $\Delta G_1^{00} \neq \Delta G_2^{00}$ and higher than the intrinsic thermodynamic coercive field $E_c$ [20, 42] in thin BaTiO$_3$ films in the actual range of temperatures and thickness [see **Figs S2(a)** in **Suppl.Mat.**]. The field prevents the domain formation in the films of thickness less than 40 nm at the temperatures T≥300 K [**Fig. S2(b)** in **Suppl.Mat.**]. The detailed interplay between domain formation and chemical screening depending on the fabrication pathway will be analyzed elsewhere.

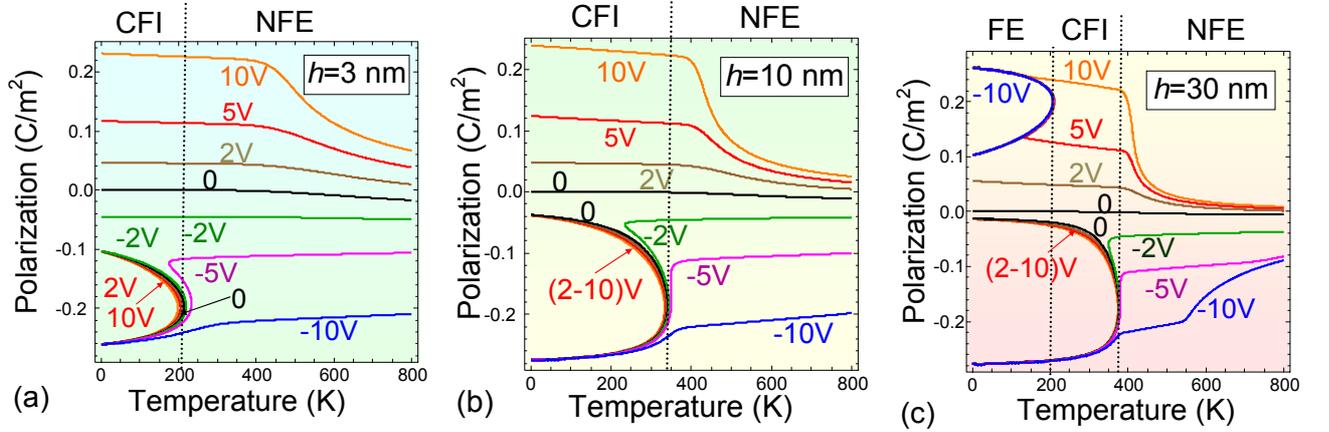

**FIGURE 2.** Temperature dependences of the average polarization calculated for different values of BaTiO$_3$ film thickness $h$=3 nm **(a),** 10 nm **(b)** and 30 nm **(c),** and gap thickness λ=0.4 nm. Each curve corresponds to one of the voltage values $U$= − 10, −5, −2, 0, 2, 5, 10 V (numbers near the curves of different colors). Vertical lines delineate non-ferroelectric ionic (**NFE**), coupled ferroelectric-ionic (**CFI**) and ferroelectric (**FE**) states. Parameters p$_{O2}$=10$^{-1}$ bar, $\Delta G_1^{00} = 1$ eV and $\Delta G_2^{00} = 0.1$ eV. Other parameters are listed in **Table SI** in **Suppl.Mat.**

We further analyze the relaxation of the average polarization $\langle P_3 \rangle$, ion charge density σ and effective piezoresponse $d_{33}$ to the equilibrium values after the voltage $U$ is switched off. These behaviors for the case of 30-nm film are shown in **Figs. 3**. Initial values of $\langle P_3 \rangle$, σ and $d_{33}$ correspond to the different voltages $U$= (−10 − +10)V. Since as a rule ionic relaxation time τ is much higher than polarization relaxation time τ$_{LK}$, we chose $\tau = 100\tau_{LK}$. Metastable states with positive polarization and piezoresponse, which have been induced by positive voltages $U$= (+1 − +10)V, rapidly relax to zero and disappear at times $\tau \approx \tau_{LK}$ [see upper curves in **Figs.3(a)** and **3(c)**]. This happens because σ is almost zero at $U>0$ [see zero horizontal line in **Fig.3(b)**].

For the case $U \leq 0$ the charge $\sigma$ is nonzero and voltage-dependent. The charge relaxation to the equilibrium value $\sigma_S \approx -0.26$ C/m$^2$ (that is equal to the spontaneous polarization of bulk BaTiO$_3$ at room temperature) starts for times $t \geq (0.5-50)\tau_{LK}$, at that the relaxation rate depends on $U$ value [compare different curves for $U \leq 0$ in **Fig.3(b)**]. The stable polar states with negative polarization and piezoresponse, which have been induced by negative and zero voltages $U = (-10 - 0)$V, demonstrate much more complex behavior in comparison with positive ones. Negative $\langle P_3 \rangle$ and $d_{33}$ induced by high negative voltage $U = -10$V at first tend to zero, until corresponding surface charge conserves for $t \ll \tau$, signifying the suppression of FE state by unscreened polarization. After the charge relaxation to the value $\sigma_S \approx -0.26$ C/m$^2$, $P_3$ and $d_{33}$ abruptly return to their equilibrium negative values $P_S \approx -0.26$ C/m$^2$ and $d_{33} \approx -26$ pm/V [see lower curves in **Figs. 3(a)** and **3(c)**]. We argue that the asymmetric multi-step relaxation between two "negative" polarizations and piezoresponces (one is the stable and another is the unstable root of the static Eq.(2a) and (3)) originates from the bistability of ionic states with respect to applied voltage. The relaxation of effective piezoresponse at lower temperature 100 K, for which both positive and negative states appear, is shown in **Fig. S4** from **Suppl.Mat.** The relaxation is complicated, nonmonotonic, and corresponds to voltage-dependent relaxation rates of polarization and surface charge. The nontrivial relaxation behavior is a characteristic feature of CFI states. This process is slightly pressure-dependent and strongly dependent on the surface ionic species (ion or vacancy) formation energies $\Delta G_i^{00}$ [see **Fig. S3** in **Suppl.Mat.**].

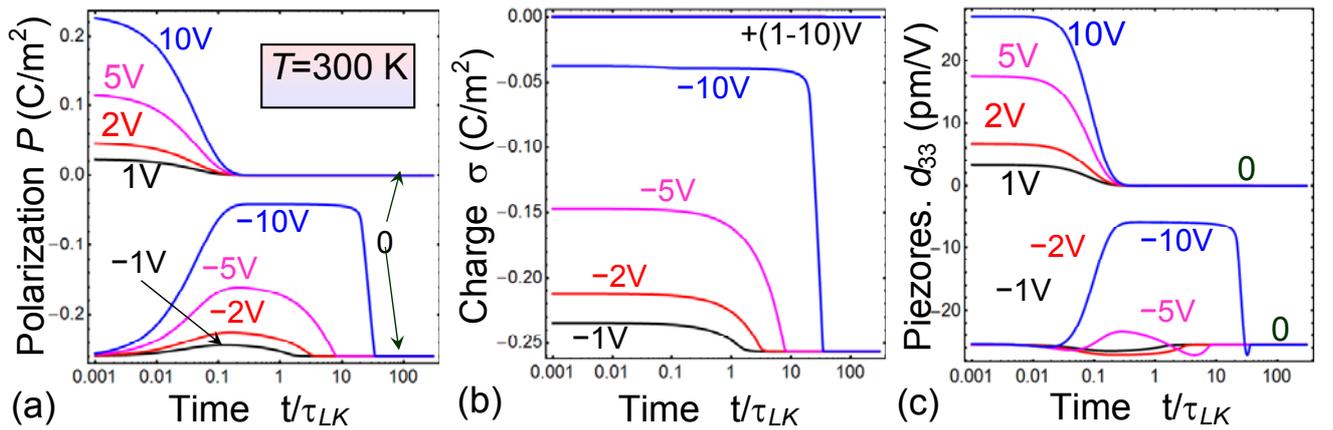

**FIGURE 3.** Average polarization $\langle P_3 \rangle$ (a), ion screening charge density $\sigma$ (b) and effective piezoresponse $d_{33}$ (c) relaxation to the equilibrium values after the applied voltage $U$ is switched off. Initial values of $\langle P_3 \rangle$, $\sigma$ and $d_{33}$ correspond to $U= -10, -5, -2, -1, 1, 2, 5$ and $10$V (specified by numbers near the curves). Film thickness $h$=30 nm, $\tau = 100\tau_{LK}$, $T= 300$ K. Other parameters are the same as in **Fig .2.**

Finally, we analyze the hysteresis dynamics of effective piezoresponse on periodic applied voltage, $U(t) = U\sin(\omega t)$, calculated for different values of dimensionless frequency $\omega\tau_{LK}$. Typical piezoresponse loops corresponding to the BaTiO$_3$ film thickness $h = (3 - 100)$ nm are shown as different curves in **Figs. 4(a)-(c)**. At low frequencies $\omega\tau_{LK} = 0.01$ the loops demonstrate multiple peculiarities (strong horizontal shift, sharp and diffuse maximums, multiple steps), which are typical to CFI states [**Fig. 4(a)**]. The principal difference between the loops calculated for $h \geq 30$ nm and $h \leq 10$ nm are clearly seen. For higher frequency $\omega\tau_{LK} = 1$ the shift increases, and the loops for the thinnest films disappears, as well as shape of the loops for thicker films becomes similar and diffuse [**Fig. 4(b)**]. At high frequency $\omega\tau_{LK} = 10$ the strong "blowing" of the loops occurs since the ion charge relaxation does not take place [**Fig. 4(c)**]. Time sweeps of effective piezoresponse over a period corresponding to the loops in **Fig.4** are shown in **Figs. S7** from **Suppl.Mat.** Hence the piezoresponses of FE and CFI states are inseparable in thermodynamic equilibrium, and become separable in high frequency dynamic mode due to the several orders of magnitude difference in the relaxation times of ferroelectric polarization and surface ions.

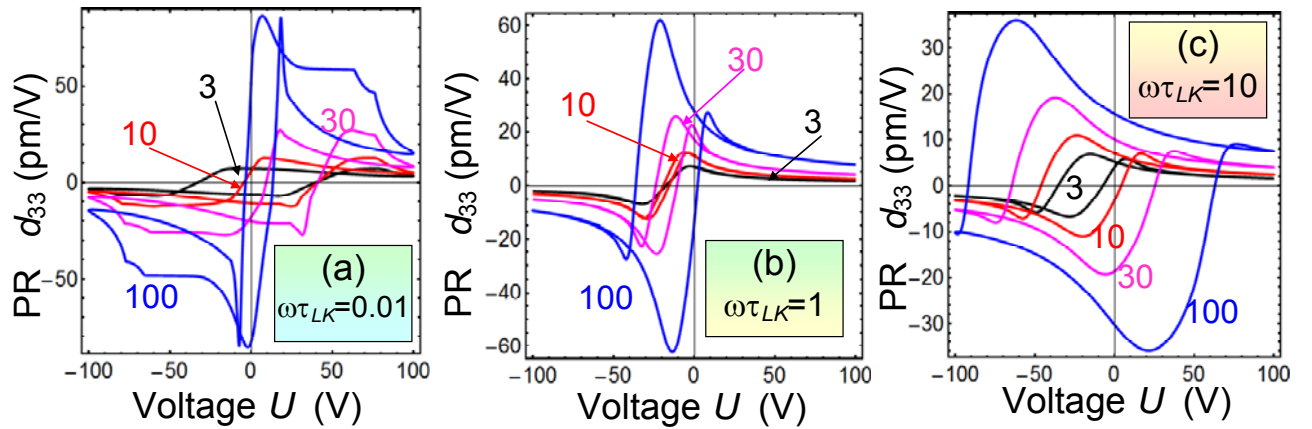

**FIGURE 4.** Effective piezoresponse (**PR**) calculated at different values of dimensionless frequency $\omega\tau_{LK}$ for several thicknesses of the film (3, 10, 30, 100) nm specified by numbers near the curves. $\tau = 100\tau_{LK}$, T=300 K. Other parameters are the same as in **Fig.2**.

To summarize, we analyzed the response of the coupled ferro-ionic states on the step wise change in electric boundary conditions, similar to the PFM imaging and spectroscopy experiment. The evolution of electromechanical response and corresponding polar states are analytically described. The responses are found to be inseparable in thermodynamic equilibrium and at low frequencies of applied voltage, and become separable in high frequency dynamic mode due to the several orders of magnitude difference in the relaxation times of ferroelectric polarization and surface ions. These studies provide a an insight into mesoscopic properties of ferroelectric thin films, whose surface is exposed to different kinds of electrochemically active gaseous surrounding.

**Supplementary Materials**, which include calculations details and auxiliary figures [43]

**Acknowledgements.** S.V.K. research is sponsored by the Division of Materials Sciences and Engineering, BES, US DOE. A portion of this research (S.V.K.) was conducted at the Center for Nanophase Materials Sciences, which is a DOE Office of Science User Facility. A.N.M., E.A.E. and N.V.M. acknowledge the NAS of Ukraine.

This manuscript has been authored by UT-Battelle, LLC, under Contract No. DE-AC0500OR22725 with the U.S. Department of Energy. The United States Government retains and the publisher, by accepting the article for publication, acknowledges that the United States Government retains a non-exclusive, paid-up, irrevocable, world-wide license to publish or reproduce the published form of this manuscript, or allow others to do so, for the United States Government purposes. The Department of Energy will provide public access to these results of federally sponsored research in accordance with the DOE Public Access Plan (http://energy.gov/downloads/doe-public-access-plan).

# 43 SUPPELEMENTARY MATERIALS
# APPENDIX A.
## Basic equations and boundary conditions

Quasi-static electric field inside the ferroelectric film is defined via electric potential as $E_3 = -\partial \varphi_f / \partial x_3$, where the potential $\varphi_f$ satisfies a conventional electrostatic equation (see below).

The system of electrostatic equations for each of the medium (gap and ferroelectric film) acquires the form:

$$\Delta \varphi_d = 0, \quad \text{(inside the gap } -\lambda \leq z \leq 0\text{)} \quad \text{(S.1a)}$$

$$\left( \varepsilon_{33}^b \frac{\partial^2}{\partial z^2} + \varepsilon_{11}^f \Delta_\perp \right) \varphi_f = \frac{1}{\varepsilon_0} \frac{\partial P_3^f}{\partial z}, \quad \text{(inside the ferroelectric film } 0 < z < h\text{)} \quad \text{(S.1b)}$$

3D-Laplace operator is $\Delta$, 2D-Laplace operator is $\Delta_\perp$.

Boundary conditions (**BCs**) to the system (S.1) are the equivalence of the electric potential to the applied voltage $U$ at the top electrode (or SPM tip apex modeled by the flat region $z = -\lambda$) and the equivalence of the normal component of electric displacements to the ion surface charge density $\sigma(\varphi)$ at $z = 0$; the continuity of the electric potential and normal component of displacements $D_3 = \varepsilon_0 E_3 + P_3$ and $D_3 = \varepsilon_0 \varepsilon_d E_3$ at gap - ferroelectric interface $z = 0$; and zero potential at the bottom conducting electrode $z = h$ [see **Figs.S1**]. Hence the **BCs** have the form:

$$\varphi_d \big|_{z=-\lambda} = U, \quad \left( \varphi_d - \varphi_f \right) \big|_{z=0} = 0, \quad \varphi_f \big|_{z=h} = 0, \quad \text{(S.2a)}$$

$$\left( \varepsilon_0 \varepsilon_d \frac{\partial \varphi_d}{\partial z} + P_3^f - \varepsilon_0 \varepsilon_{33}^b \frac{\partial \varphi_f}{\partial z} - \sigma \right) \bigg|_{z=0} = 0. \quad \text{(S.2b)}$$

The polarization components of uniaxial ferroelectric film depend on the inner field $E$ as $P_1 = \varepsilon_0 (\varepsilon_{11}^f - 1) E_1$ and $P_2 = \varepsilon_0 (\varepsilon_{11}^f - 1) E_2$ and $P_3(\mathbf{r}, E_3) = P_3^f(\mathbf{r}, E_3) + \varepsilon_0 (\varepsilon_{33}^b - 1) E_3$. The relative dielectric permittivity $\varepsilon_{33}^f$ related with the ferroelectric polarization $P_3$ via the soft mode. The temporal evolution and spatial distribution of the ferroelectric polarization $P_3^f$ (further abbreviated as $P_3$) is given by the time-dependent LGD equation:

$$\Gamma \frac{\partial P_3}{\partial t} + a_3 P_3 + a_{33} P_3^3 + a_{333} P_3^5 - g_{33} \frac{\partial^2 P_3}{\partial z^2} = E_3, \quad \text{(S.3a)}$$

with the **BCs** of the third kind at the film surfaces $z = 0$ and $z = h$:

$$\left( P_3 - \Lambda_- \frac{\partial P_3}{\partial z} \right) \bigg|_{z=0} = 0, \quad \left( P_3 + \Lambda_+ \frac{\partial P_3}{\partial z} \right) \bigg|_{z=h} = 0, \quad \text{(S.3b)}$$

which include extrapolation lengths $\Lambda_+$ and $\Lambda_-$.

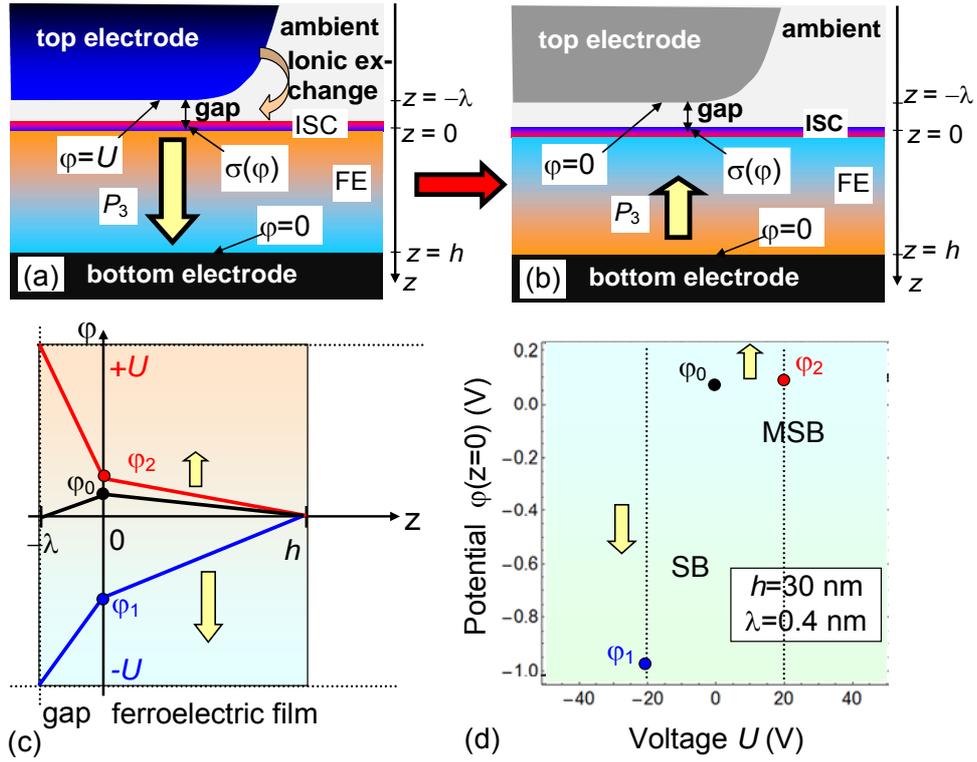

**FIGURE S1**. **(a-b)** Layout of the considered system, consisting of electron conducting bottom electrode, ferroelectric (FE) film, ion surface charge (ISC), ultra-thin gap allowing direct ion exchange with environment, and the tip electrode (from bottom to the top). **(a)** Voltage $U$ is applied to the tip electrode. **(b)** The voltage is switched off. **(c)** Schematics of electric potential z-profiles for positive (+U) and negative (-U) voltages applied to the tip (red and blue broken lines). Z-profile of potential relaxed to the equilibrium after the voltage is switched off is shown by the broken line. **(d)** Dependence of the electric potential $\varphi$ at the interface $z = 0$ between ferroelectric film and ion charge layer on applied voltage. Stable (SB) and metastable (MSB) branches are shown. The arrows indicate polarization direction. BaTiO$_3$ film thickness $h = 30$ nm and gap thickness $\lambda = 0.4$ nm, pressure $p_{O2} = 10^{-1}$ bar. Other parameters are listed in **Table SI.**

**TABLE SI. Description of physical variables and their numerical values**

| Description of main physical quantities used in Eqs.(S.1)-(S.3) | Designation and dimensionality | Value for a structure BaTiO$_3$ film / ion charge / gap / tip |
|---|---|---|
| Polarization of ferroelectric along polar axis Z | $P_3$ (C/m$^2$) | variable (0.26 for a bulk material) |
| Electric field | $E_3$ (V/m) | variable |
| Electrostatic potentials of dielectric gap and ferroelectric film | $\varphi_d$ and $\varphi_f$ (V) | variables |
| Electric voltage on the tip | $U$ (V) | variable |
| Coefficient of LGD functional | $a_3 = \alpha_T(T - T_C)$ (C$^{-2}$ J m) | T-dependent variable |
| Dielectric stiffness | $\alpha_T$ ($\times 10^5$ C$^{-2}$·J·m/K) | 6.68 |
| Curie temperature | $T_C$ (K) | 381 |

| Coefficient of LGD functional | $a_{33}$ ($\times 10^9$ J C$^{-4}$·m$^5$) | $-8.18 + 0.01876 \times T$ |
|---|---|---|
| Coefficient of LGD functional | $a_{333}$ ($\times 10^{11}$ J C$^{-6}$·m$^9$) | $1.467 - 0.00331 T$ |
| Gradient coefficient | $g_{33}$ ($\times 10^{-10}$ m/F) | (0.5-5) |
| Kinetic coefficient | $\Gamma$ (s× C$^{-2}$ J m) | rather small |
| Landau-Khalatnikov relaxation time | $\tau_K$ (s) | $10^{-11} - 10^{-13}$ (far from Tc) |
| Thickness of ferroelectric layer | $h$ (nm) | variable 3 – 500 |
| Background permittivity of ferroelectric | $\varepsilon_{33}^b$ (dimensionless) | 10 |
| Extrapolation lengths | $\Lambda_-$, $\Lambda_+$ (angstroms) | $\Lambda_- = 1$ Å, $\Lambda_+ = 2$ Å |
| Surface charge density | $\sigma(\varphi,t)$ (C/m$^2$) | variable |
| Equilibrium surface charge density | $\sigma_0(\varphi)$ (C/m$^2$) | variable |
| Occupation degree of surface ions | $\theta_i$ (dimensionless) | variable |
| Oxygen partial pressure | $p_{O2}$ (bar) | variable |
| Surface charge relaxation time | $\tau$ (s) | >> Landau-Khalatnikov time |
| Thickness of dielectric gap | $\lambda$ (nm) | 0.4 |
| Permittivity of the dielectric gap | $\varepsilon_d$ (dimensionless) | 1 |
| Universal dielectric constant | $\varepsilon_0$ (F/m) | $8.85 \times 10^{-12}$ |
| Electron charge | $e$ (C) | $1.6 \times 10^{-19}$ |
| Ionization degree of the surface ions | $Z_i$ (dimensionless) | $Z_1 = +2$, $Z_2 = -2$ |
| Number of surface ions created per oxygen molecule | $n_i$ (dimensionless) | $n_1 = 2$, $n_2 = -2$ |
| Inverse concentration of the surface ions at saturation | $A_i$ (m$^2$) | $A_1 = A_2 = 2 \times 10^{-19}$ |
| Surface defect/ion formation energy | $\Delta G_i^{00}$ (eV) | $\Delta G_1^{00} = 1$, $\Delta G_2^{00} = 0.1$ |
| Gradient-correlation size effect lengthscale | $L_g = \dfrac{g_{33}(\Lambda_+ + \Lambda_-)}{\alpha_T T_C \Lambda_+ \Lambda_-}$ (nm) | 5.92 for $g_{33}=1$ |
| Depolarizing size effect lengthscale | $L_d \approx \dfrac{L_S}{\varepsilon_0 \alpha_T T_C}\left(1 + L_S \dfrac{\varepsilon_d}{\lambda}\right)^{-1}$ (nm) | 4.5 (calculated at $L_S = 0.1$ Å) |
| Dielectric gap field lengthscale at $\sigma = 0$ | $L_\lambda = \dfrac{\lambda}{\varepsilon_0 \varepsilon_d \alpha_T T_C}$ (nm) | 178 (calculated at $\lambda = 0.4$ nm) |
| Film critical thickness | $h_{cr}(T) \approx \dfrac{T_C(L_g + L_D)}{(T_C - T)}$ (nm) | 49.4 nm (calculated at T = 300K) |

## APPENDIX B.

## B.1. Single-domain state stability

Despite the ultra-thin (0.4 nm) gap between the sluggish ion charge layer and the top electrode is present in the geometry of this work, the situation with domain appearance is completely different due to the strong self-polarizing role of effective electric field $E_{eff}(U = 0, \sigma) = \dfrac{\lambda \sigma}{\varepsilon_0(\varepsilon_d h + \lambda \varepsilon_{33}^b)}$ produced by nonzero ion charge $\sigma$ at U = 0. The field, being inversely proportional to the film thickness (and so it is predicted to be the highest for thin films), is nonzero and rather high for the case of different ion formation energies $\Delta G_1^{00} \neq \Delta G_2^{00}$ (namely, we put $\Delta G_1^{00} = 1\,\text{eV}$ and $\Delta G_2^{00} = 0.1$ eV). The field $E_{eff}(0,\sigma)$ becomes higher than the thermodynamic coercive field $E_c$ in thin BTO films in the actual range of temperatures and thickness [see **Figs S2(a)-(b)** below]. So that it indeed self-polarizes thin BTO films covered with ion layer and prevents the domain formation for the films of thickness less than (10-20) nm at arbitrary temperatures; they may appear only for 30-nm films at temperatures less than 270 K [**Fig. S2(a)**]. At the temperatures T ≥ 300 K the domains are absent for thickness less than 40 nm (we additionally checked the last statement numerically) [**Fig. S2(b)**].

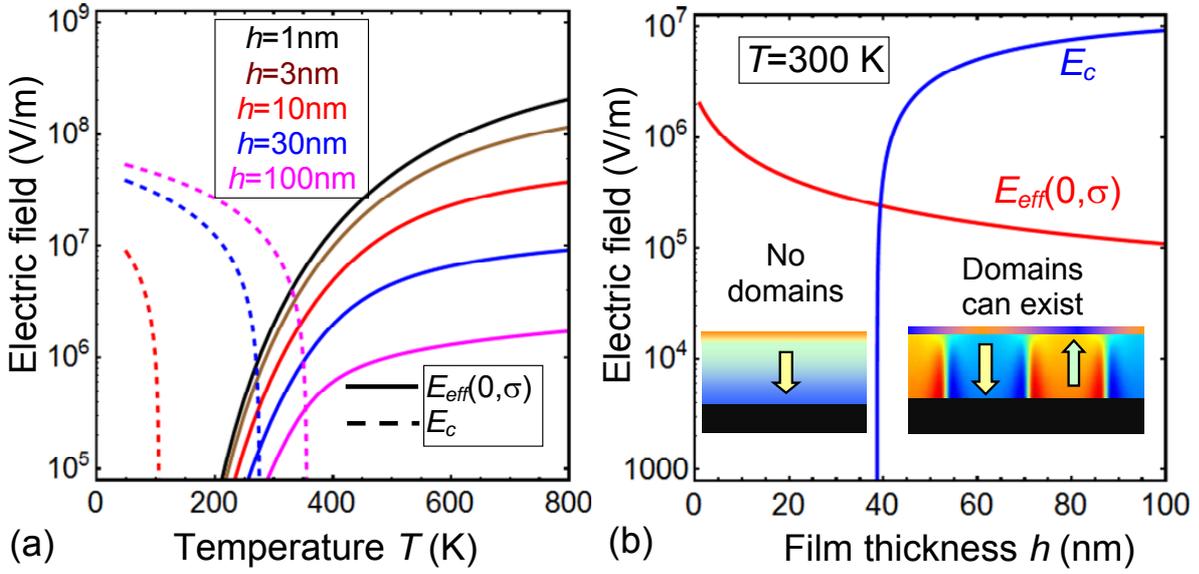

**Figure S2.** (a) Temperature dependences of effective electric field $E_{eff}(U=0,\sigma)$ (solid curves) and thermodynamic coercive field $E_c$ (dashed curves) calculated for different thickness $h$ (from 1 to 100 nm) of BTO film. (b) Thickness dependences of effective electric field $E_{eff}(U=0,\sigma)$ (red curve) and thermodynamic coercive field $E_c$ (blue curve) calculated at room temperature. Detailed parameters are described in Table SI.

Note, that the thermodynamic **intrinsic** coercive field $E_c$ was found directly from Eqs.(1)-(3) namely the field is $E_c = \dfrac{2}{3\sqrt{3}}\sqrt{-\dfrac{a_R^3}{a_{33}}}$ for the ferroelectrics with the second order phase transition and

$$E_c = \frac{2}{5}\left(2a_{333} + \sqrt{9a_{33}^2 - 20a_R a_{333}}\right)\left(\frac{2a_R}{-3a_{33} - \sqrt{9a_{33}^2 - 20a_R a_{333}}}\right)^{3/2}$$

for the ferroelectrics with the first order phase transition, respectively. The coefficient

$a_R \approx \alpha_T(T_C - T) + \dfrac{g_{33}}{h}\left(\dfrac{1}{\Lambda_+} + \dfrac{1}{\Lambda_-}\right) + \dfrac{L_D}{\varepsilon_0(\varepsilon_d h + L_D \varepsilon_{33}^b)}$. The expression accounts for both the gradient and gap effect. The depolarization field length is $L_D = \dfrac{L_S}{\varepsilon_0 \alpha_T T_C}\left(1 + L_S \dfrac{\varepsilon_d}{\lambda}\right)^{-1} \approx \dfrac{L_S}{\varepsilon_0 \alpha_T T_C}$, where $L_S = 0.1$ Å is the surface screening length of the ion charge.

Also the domain formation does not affect on the "sharp" boundary between NFE and CFI states, it can only diffuses the boundary between FE and CFI states for the films thicker than 20 nm (since the domains are absent for thinner films).

We made additional calculations to compare the three electrostatic free energies of the (either PTO or BTO) ultra-thin film at zero applied voltage $U=0$, namely with domain structure but without top ion layer (case 1); with domain structure and top ion layer that provides effective screening (case 2); without domain structure and with top ion layer that provides effective screening (case 3). For all cases the film is in a perfect electric contact with electron-conducting bottom electrode. Appeared that the term $-\int_S dxdy \int_0^h dz (P_3 E_3)$ in the LGD free energy ($P_3$ is polarization, $E_3$ is internal electric field) is in fact a positive depolarization field energy in the case 1. It becomes a negative quantity in the case 2, and is very small in comparison with the case 1. The domain walls surface energy is positive in the case 2. The term becomes a noticeable negative quantity in the case 3, due to the effective "ionic" field, i.e. $-\int dz (P_3 E_3) \sim \dfrac{h}{2}\langle P_3 \rangle E_{eff}$ and in the single-domain case 3 the positive domain walls energy is minimal. Hence the case 3 is the most energetically preferable as anticipated if only $E_{eff}(U=0, \sigma)$ is rather high.

**B.2. Temperature dependences of the equilibrium average polarization and ion charge**

Temperature dependences of the equilibrium **single-domain** average polarization $\langle P_3(T) \rangle$ and surface ionic screening charge density calculated for different values of applied voltage are shown in **Figs.S3.**

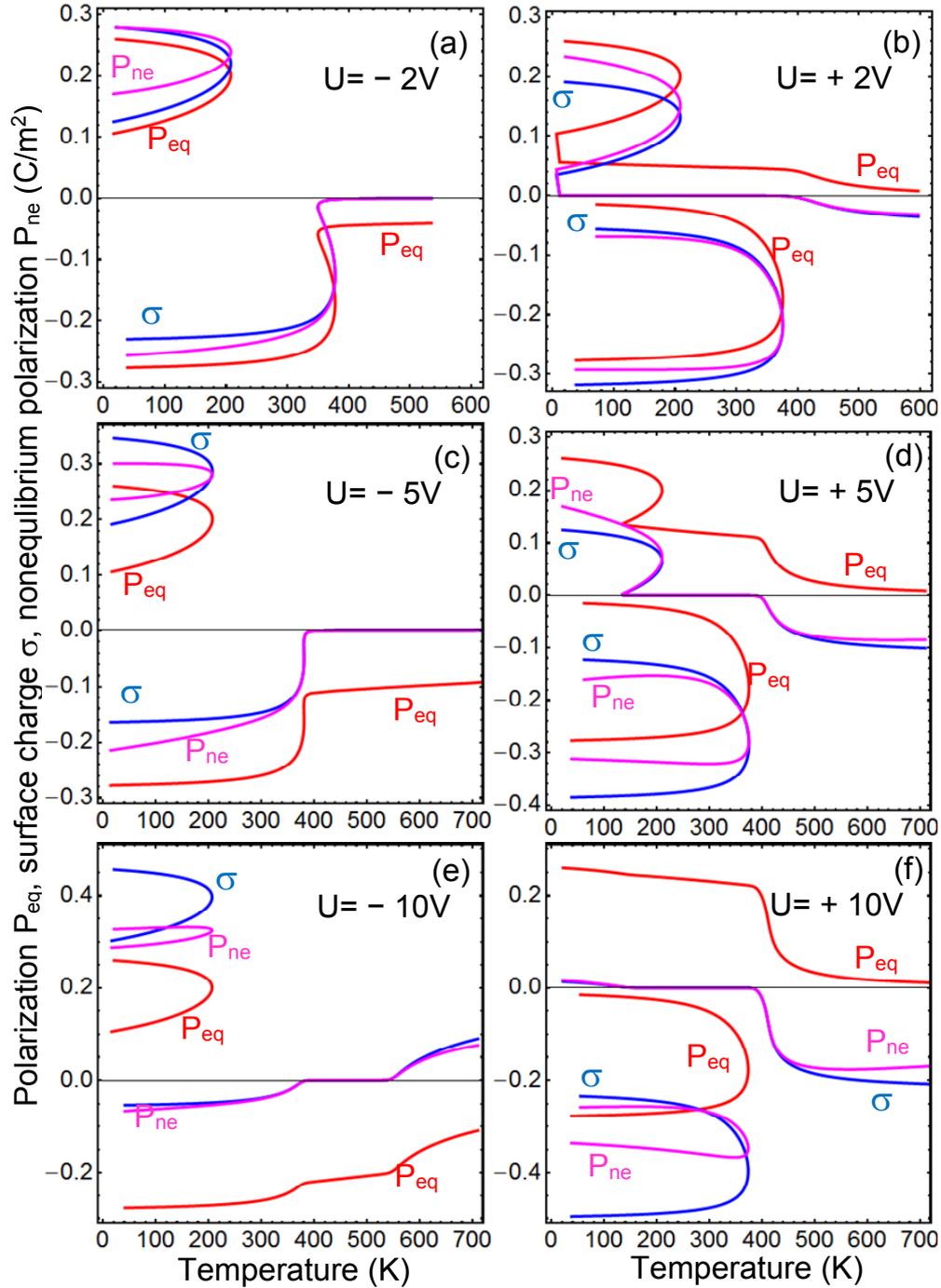

**FIGURE S3.** Temperature dependences of the equilibrium average polarization (red curves) and ionic screening charge (blue curves) calculated for different values of applied voltage $U = (-2, 2, -5, 5, -10, 10)$V (plots **(a)**, **(b)**, **(c)**, **(d)**, **(e)** and **(f)** respectively), and non-equilibrium polarization (magenta curves) calculated in the case of instant removal of tip voltage ($U = 0$) and very sluggish relaxation of the ion charge. BaTiO$_3$ film thickness $h = 30$ nm. Pressure $p_{O2} = 10^{-1}$ bar. Other parameters are the same as in **Fig.S2.**

## APPENDIX C
### Relaxation of polarization, ion charge and effective piezoresponse

The relaxation of polarization, ion charge and effective piezoresponse at lower temperature 100 K is shown in **Figs.S4.** The relaxation is complicated, nonmonotonic, and corresponds to voltage-dependent relaxation rates of polarization and surface charge for which both positive and negative states exists. The system piezoresponse relaxes to the values $d_{33} = +20$ pm/V for positive poling voltages $U = (+1 - +10)$ V, or to the values $d_{33} = -17$ pm/V for negative poling voltages $U = -(1 - 10)$ V.

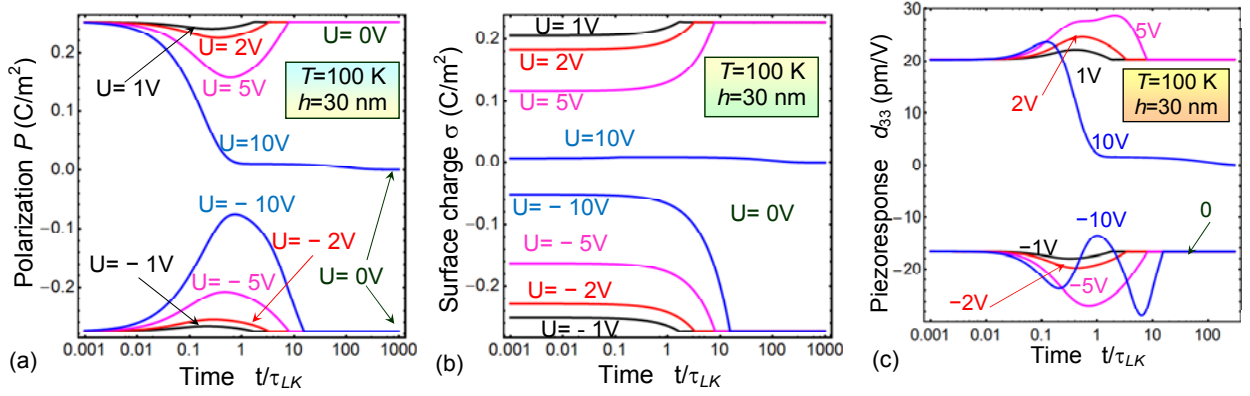

**FIGURE S4.** Average polarization $\langle P_3 \rangle$ (a), ionic screening charge density σ (b) and effective piezoresponse $d_{33}$ (c) relaxation to the equilibrium values after the applied voltage $U$ is switched off. Initial values of $\langle P_3 \rangle$, σ and $d_{33}$ correspond to the different values of $U = -10, -5, -2, -1, 1, 2, 5$ and $10$ V (specified by numbers near the curves). Film thickness is $h=30$ nm, $\tau = 100\tau_{LK}$, temperature $T = 100$ K. Other parameters are the same as in **Fig.S2.**

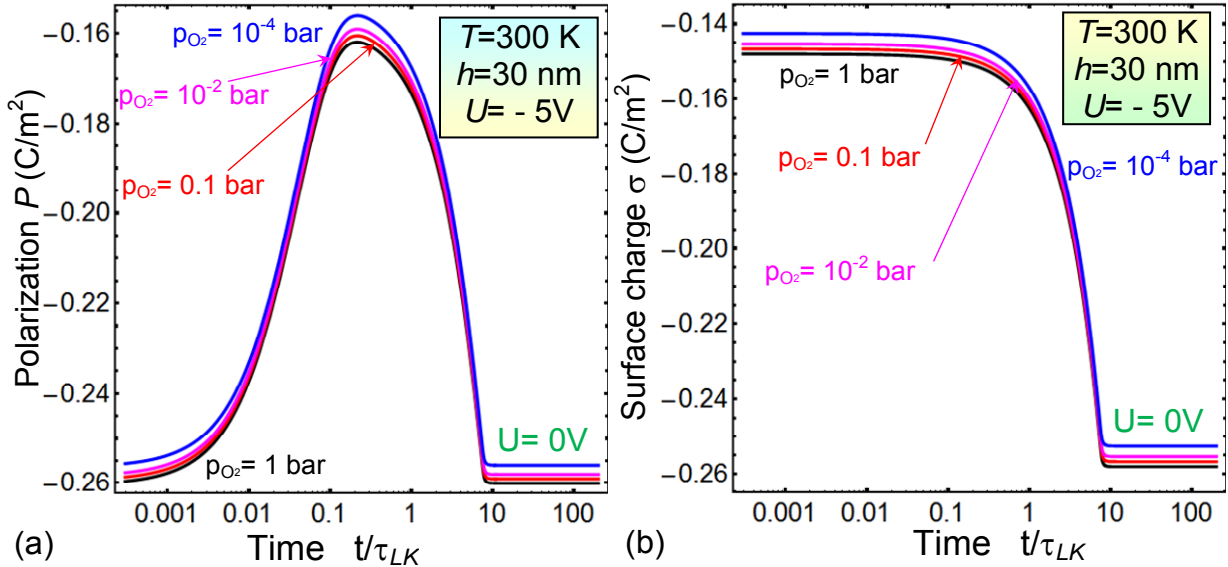

**FIGURE S5.** Relaxation with time of the equilibrium (a) average polarization and (b) ionic screening charge to the equilibrium values when the voltage is switched off, U = 0V. Initial values of polarization and charge correspond to the different values of oxygen partial pressure $p_{O2}$ specified near the curves and applied voltage $U = -5$ V. Film thickness h=30 nm, temperature T= 300 K.

## APPENDIX D. Effective piezoresponse hysteresis loops

Finally, we analyze the hysteresis dynamics of effective piezoresponse on periodic applied voltage, $U(t) = U\sin(\omega t)$. Effective piezoresponse of BaTiO$_3$ films was calculated at 300K for different dimensionless frequency $w = \omega\tau_{LK}$ and film thickness. Typical piezoresponse loops corresponding to the film thickness $h$=(3 - 100) nm are shown as different curves in **Figs. S6(a)-(d)**. At low frequencies $w = 0.01$ (corresponding to $\omega\tau = 1$) piezoresponse loops demonstrate multiple peculiarities (strong horizontal shift, sharp and diffuse maximums, multiple steps) [**Fig. S6(a)**], which are typical to CFI states. The frequency increase to 0.1 smears and equalizes all maximums and leads to the steps disappearance, but the principal differences between the piezoresponse loops for the films of thickness $h > h_{cr}(T)$ (100 nm and 30 nm) and $h > h_{cr}(T)$ (3 nm and 10 nm) are still clearly seen [**Fig. S6(b)**]. This take place because the ionic relaxation time τ becomes much smaller than the period of applied voltage $2\pi/\omega$. For middle frequency $w = 1$ (corresponding to $\omega\tau = 100$) the shift increases, and the loops for the thinnest films disappears, as well as shape of the loops for thicker films becomes similar and diffuse [**Fig. S6(c)**]. At very high frequency $w = 10$ (corresponding to $\omega\tau = 10^3$) the strong "blowing" of the loops occurs for all films since the ion charge relaxation does not take place [**Fig. S6(d)**].

Time sweeps of effective piezoresponse over a period corresponding to the loops in **Fig.S6** are shown in **Figs. S7(a)-(d).** The piezoresponse peaks correspond to polarization reversal. The sharp and high peaks correspond to thicker films; diffuse and low peaks are characteristic for thin ones. The smearing of the piezoresponse peaks and the disappearance of the secondary peculiarities, which are inherent to CFI states, occurs in thinner films with the frequency increase.

Hence the piezoelectric responses of ferroelectric and ionic states are inseparable in thermodynamic equilibrium, and become separable in high frequency dynamic mode due to the several orders of magnitude difference in the relaxation times of ferroelectric polarization and surface ions.

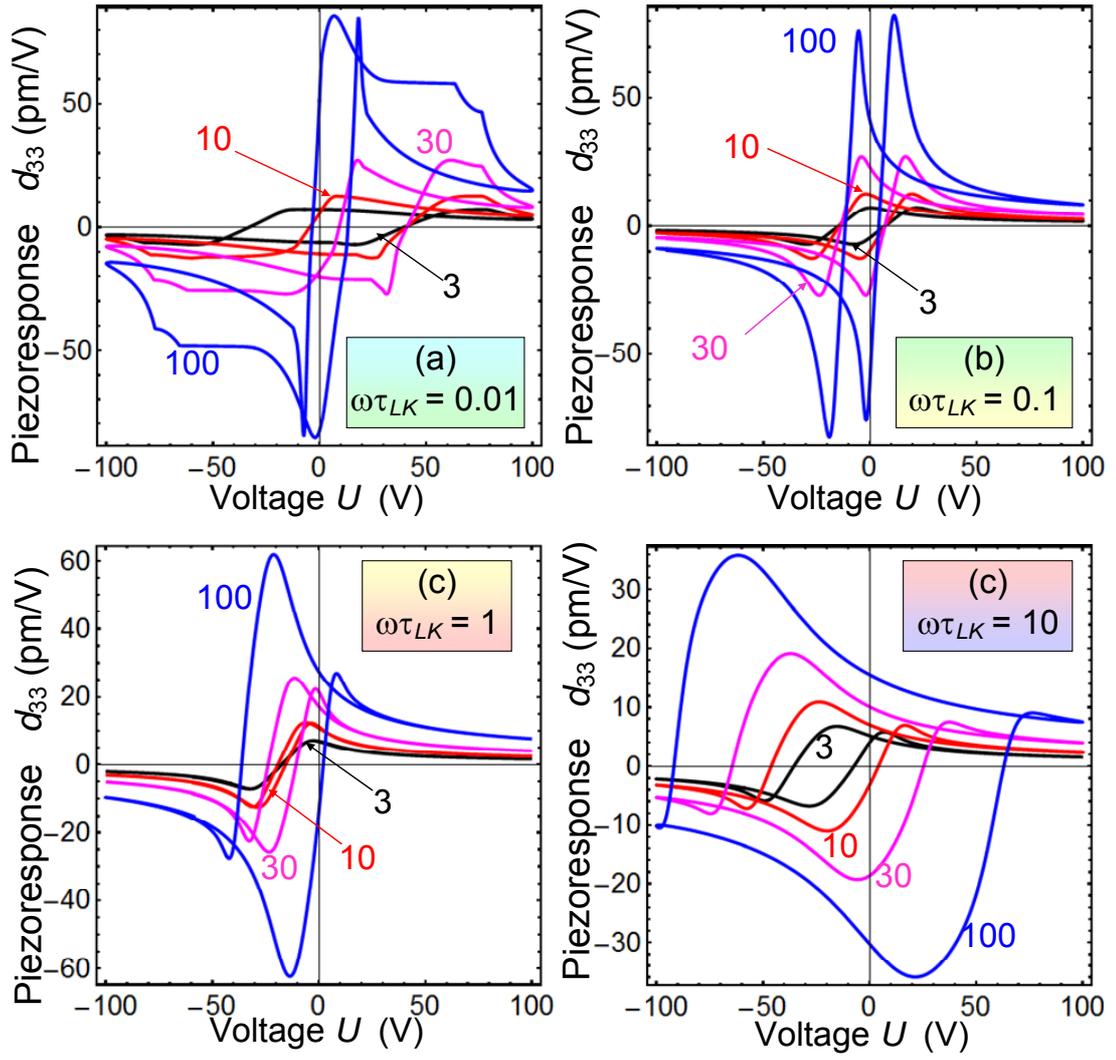

**FIGURE S6.** Effective piezoresponse calculated at different dimensionless frequency $\omega\tau_{LK}$ for different thickness of the film $h$ = (3, 10, 30, 100) nm specified by numbers near the curves. $\tau = 100\tau_{LK}$, T = 300 K. Other parameters are the same as in **Fig.2**, main text.

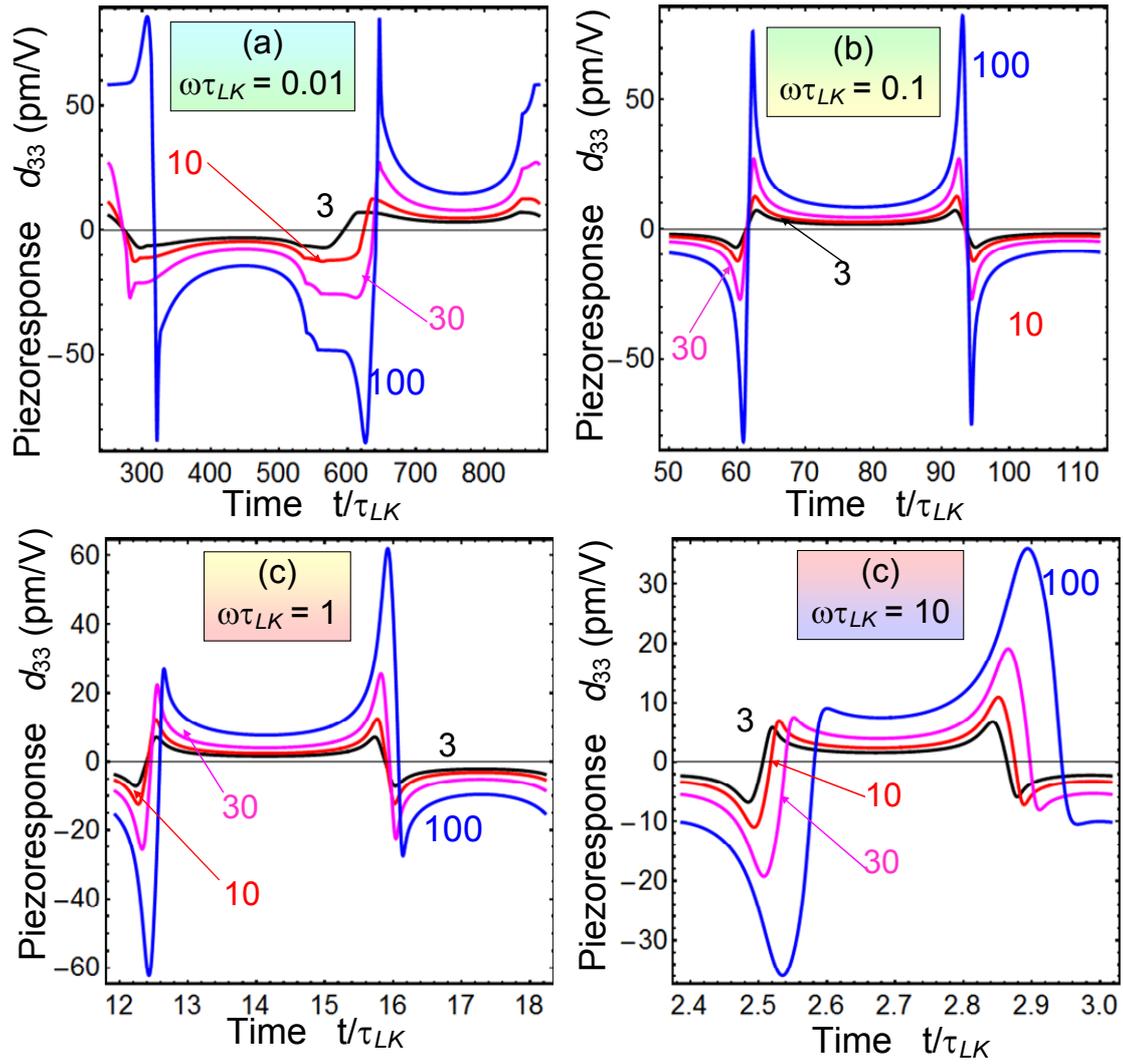

**FIGURE S7.** Time sweeps of effective piezoresponse calculated at different dimensionless frequency $\omega\tau_{LK}$ for different thickness of the film $h = (3, 10, 30, 100)$ nm specified by numbers near the curves. $\tau = 100\tau_{LK}$, T = 300 K. Other parameters are the same as in **Fig.2**, main text.